\documentclass[manuscript,nonacm]{acmart}
\AtBeginDocument{%
  }

\setcopyright{cc} 
\copyrightyear{2026}
\setcctype[4.0]{by}

\begin{document}

\title{Time to Get Closer: Longing for Care Ethics Under the Neoliberal Logic of Public Services}

\author{Rūta Šerpytytė}
\email{ruta.serpytyte@tuni.fi}
\orcid{0000-0002-7647-761X}
\affiliation{
  \institution{Tampere University}
  \city{Tampere}
  \country{Finland}
}


\begin{abstract}
The fields of HCI and Participatory design have been turning to care ethics as a suitable ethos to approach current polycrisis with. Similar calls for relationality can be witnessed in public administration research and practice, albeit its current logic being built on privatisation and marketisation of services, managerialism and customer-focus; all of which are challenging to combine with care ethics. In this paper I use collaging technique to visually reflect on new ways for public services to adopt and (care-fully) scale participatory design approaches, and how do feminist care ethics fit in the design of public services, where there is a strong presence of neoliberalism.
\end{abstract}






\maketitle
\small This work is part of the \emph{Proceedings of CHIdeology Workshop 2026}, held on Wednesday, 15 April, in Barcelona, Spain, at CHI’26.

\begin{figure}
    \centering

    \includegraphics[width=0.8\textwidth]{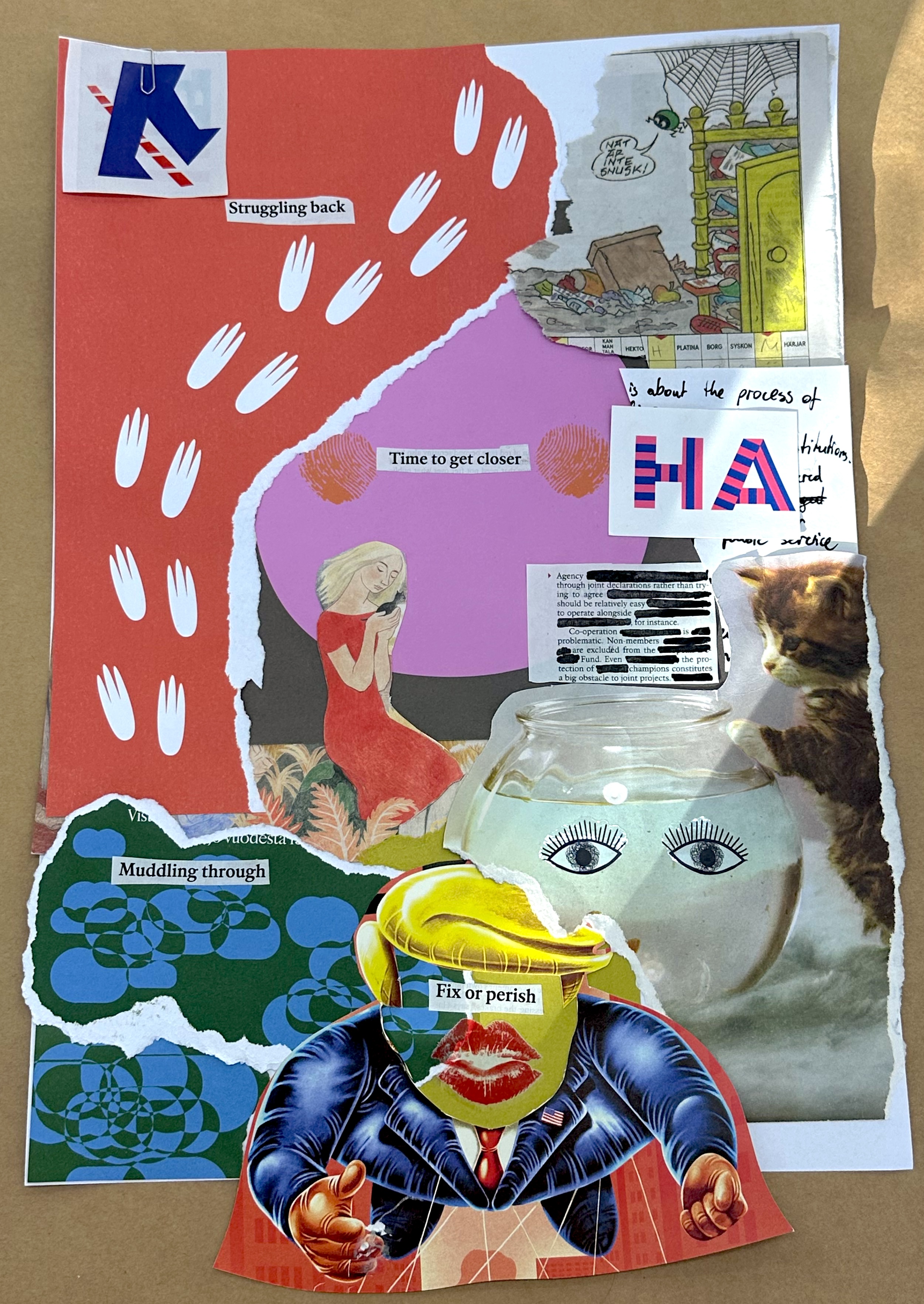}
    \caption{A collage illustrating the process of moving towards care ethics in public service development.}
    \Description{A collage which contains an illustration of D. Trump's silhouette with words "Fix or perish", abstract intertwined shapes with words "Muddling through", Moomin footsteps with words "Struggling back", a graphic of legs stepping over a line, a comic of a web, spider, and dirty floor, big letters "HA", a "blackout poem", and a picture of a cat looking into an aquarium. In the centre, there's an illustration of a woman gently holding a kitten, the pink background has two fingerprints with text "Time to get closer".}
    \label{collage}
  \end{figure}

\section{Introduction}

Participatory design (PD) emerged in the 70s when researchers built partnerships with trade unions to ensure workers are included in the design of new technologies. However, with the rise of neoliberalism, increasing globalisation, privatisation of the public sector, and digitalisation, the field has disengaged from the political and shifted toward "small" community engagements \cite{huybrechts_institutioning_2017}. Noting how excluding institutions in PD can perpetuate neoliberal ideals, scholars have been calling for re-politicising PD \cite{bodker_participatory_2018, smith_contemporary_2025}. This also means increasing PD's scale and scope, engaging with public institutions, operating in more complex value systems and working with different ideologies instead of ignoring them.

Care ethics follow feminist goals of redistributing power and challenging traditional (patriarchal) norms, systems, and morals. Matters of care can be understood as a three-fold notion of affects, ethics/politics and labour \cite{puig_de_la_bellacasa_matters_2017}. HCI scholarship has adopted care ethics in terms of thinking about socio-technical challenges \cite{petterson_expanding_2025, henriques_feminist_2025}. At the same time, public management scholars notice a shift towards relationality and reciprocity in the research and practice, in order to reform public services \cite{bartels_debate_2025}, which are currently grounded in New Public Management (NPM), prioritising efficiency in performance measurements. However, the wave of rapid digitalisation of public sector in order to make it smooth and efficient results in compromising its social equity measures \cite{frederickson_social_2010,blessett_social_2019}.

In my PhD research I look into the questions of scaling participatory approaches in public institutions. The work is situated in a project which aims to improve design of municipal integration services for migrant residents in a Finnish municipality. Migration is an inherently political topic, with talking points fluctuating between safety measures (protecting borders), to human rights and exposed racism. In my work, I lean towards feminist values of emancipation, social equity and community care, while I associate neoliberalism with individualism, efficiency of services, and marked-driven societies. \textit{Scaling} questions normally fall under the latter, while \textit{migrant inclusion} is better suited to be discussed in the former value system. 

It is evident that both HCI work and public sector management are affected by certain ideologies. To explore how these different strands connect (participatory design, scaling, care, neoliberalism, feminism, migration, public sector), I engaged in a collaging activity as a part of my PhD process. Collage is an inclusive yet precarious practice of making an art-efact, usually by using glue and paper, and it is suitable for interdisciplinary research, creating new ways of knowing \cite{vaughan_pieced_2005}. I present the collage as a visualisation of how different ideologies manifest in matters of design and HCI.

\section{Visual Exploration of the Topic Through Collaging}
Below I describe different sections of the collage (Fig \ref{collage}), introducing the ideas behind each choice. At the same time, I am aware that a big part of this exercise was thinking with hands, so some of the descriptive meanings were found after sticking the words and images on the paper.

\subsection{Fix or Perish}
The bottom part of the collage contains an illustration of D. Trump (\textit{The Economist}). While many, and at the same time none of the political ideologies describe policies of the current US Government, I picked the image thinking of its anti-state populist rhetoric, reducing government spending, and horrible treatment of migrant residents, connecting it to my setting. I also saw parallels (Trump being tied by strings) to NPM and its focus on performance-based metrics, applying corporate managerial logic to the public sector. In addition, I wanted to include political imagery in the composition as a reminder to avoid depoliticising of design.

\subsection{Muddling Through}
The bottom-left corner of the collage includes abstract shapes, which resonated with my methodological approach of \textit{knotworking}, which is a form of participatory infrastructuring \cite{bodker_tying_2017}. "Muddling through" is an accurate description of how such structureless approach feels like, connecting multiple actors and tying different topics and agendas together, in order to sustain the project.

\subsection{Struggling Back}
The illustration of the footprints was the first piece of paper that I picked up, as it spoke to me about scaling care-fully, which I envision as a slow step-by-step process, taking time to pause and reflect, being attentive to the needs of care-receivers, and being open to emerging conflicts ("Struggling back"). The migrant imagery at the top left corner (illustrated by Ben Hickey for \textit{The Economist}) visually connects with the footprints, in this case it's not an abstract process, but an intentional goal to cross into a new place, aspiring for belonging here.

\subsection{It's Not Filthy Here!}
On the top-right corner I stuck an illustration of a spider with a Swedish phrase "Nåt är inte snusk!", in the context of the comic roughly translated to "It's not filthy here!" (illustrated by Anna Simberg for the \textit{Hufvudstadsbladet}). The imagery reminded me of care labour, and how it usually becomes most visible in its absence. It also made me think of care as a feminist issue, as care and other affective labour are often gendered and racialised \cite{ahmed_cultural_2014}. When introducing more care ethics in the neoliberal settings of the public sector, there is a need to implement it intentionally as an active moral choice, and not base it on a socio-cultural norm, perceived as a "natural instinct" for women \cite{tronto_moral_2020}.

\subsection{The Cat and the Aquarium}
Initially, the cat in the image was looking at a goldfish in the aquarium. Ripping off the goldfish and adding eye stickers made the visual connected to introspection and pausing. Such pauses can help critically reflect on researcher's positionality, how their choices and values are enacted in their thinking and practice \cite{christensen_researchers_2024}. It can also help when acknowledging the invisible labour of care and other mundane, backstage tasks. Tronto's first care dimension is \textit{attentiveness}, which requires suspending one's own goals and ambitions in order to notice a need for care \cite{tronto_moral_2020}.

\subsection{Non-members are Excluded from the Fund}
\begin{figure}
    \centering
    \includegraphics[width=0.8\textwidth]{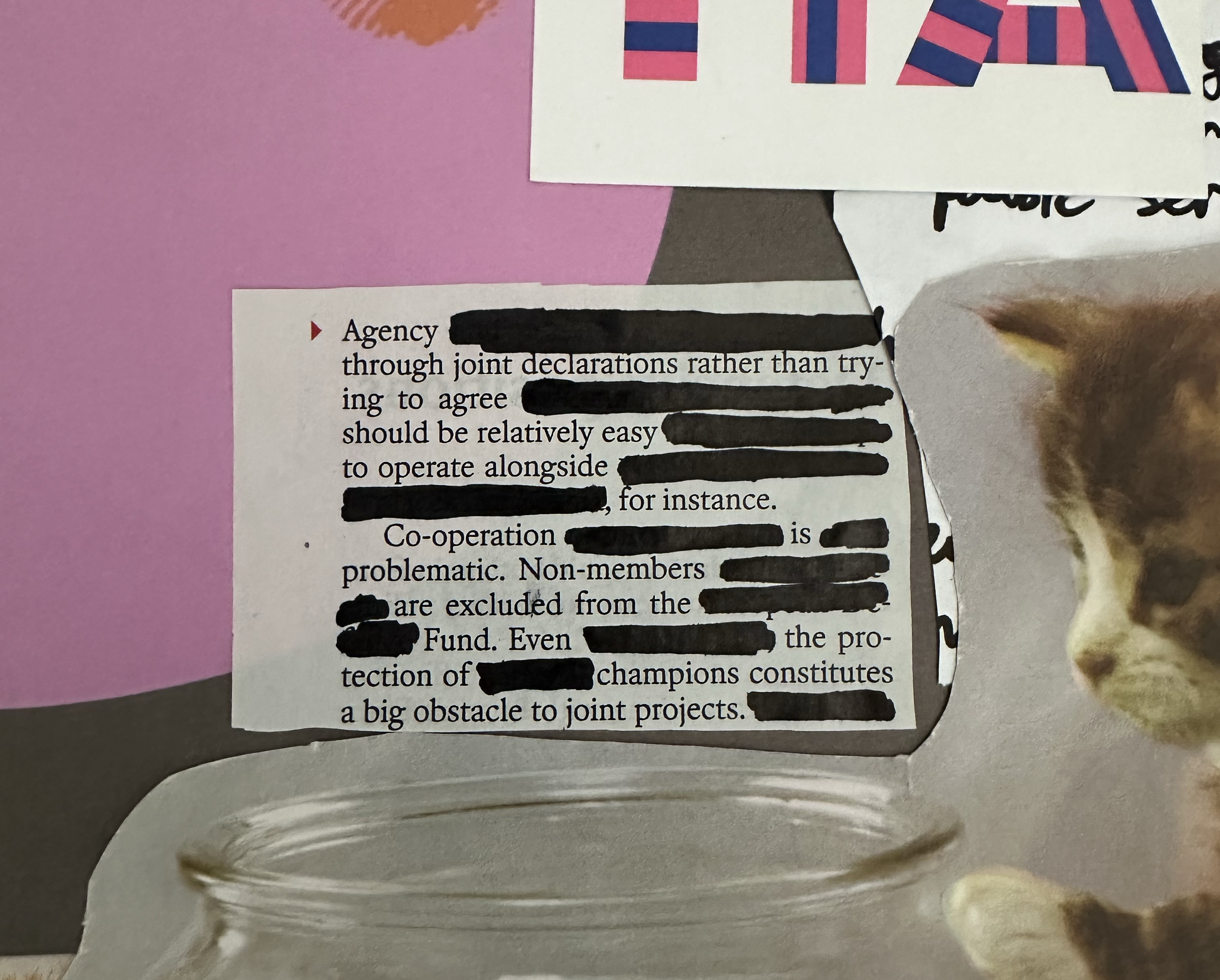}
    \caption{A piece of blackout poetry.}
    \Description{An excerpt of a newspaper texts with some parts erased. The text that's left reads: Agency through joint declarations rather than trying to agree should be relatively easy to operate alongside, for instance. Co-operation is problematic. Non-members are excluded from the Fund. Even the protection of champions constitutes a big obstacle to joint projects}
    \label{poem}
  \end{figure}

As a PD researcher, my eyes fell upon the word "co-operation". I picked up a marker and, without giving it much though, started crossing out irrelevant words, which resulted in this nonsensical meme-like poem (Fig \ref{poem}) about agonism, projectisation of PD caused by funding structures, recruiting participants and other known struggles in the field. I stuck the poem above the aquarium as an example of thoughts that can emerge from a reflexive practice.

\subsection{Time to Get Closer}
Finally, at the very centre of the image, I put "Time to get closer" between two fingerprints, as a call for implementing more care ethics in the public sector. A woman gently holding a cat (from Tove Jansson's mural "Bird Blue"), while reinforcing the stereotype of women as carers, fit nicely with this call for affection and affecting. On the right of this soft imagery, I left some of my brainstorm notes and "HA" in big letters, as a reminder for myself how PD is not a magical concept and that this idea of care in public services is not a utopia, but can only be realised as an ongoing struggle.

\section{Conclusion}
My expected contribution to "CHIdeology" workshop is two-fold. First, I want to engage in discussions on feminist versus neoliberal public services, and what the role of a PD or HCI researcher might be here. I am excited to introduce my case of striving for care-centric attitudes when scaling PD in the public sector, and share examples of working with migrant residents and NGOs. Second, I would be happy to explore collage and other artistic methods as a way to disambiguate ideologies by making them material.

\begin{acks}
The collage was created in the workshop held at Aalto University's department of Design, guided by Andrea Botero and Linnea Sjöholm. My sincere thanks for arranging this moment to think and reflect.

I acknowledge the work of other authors and creators whose work was used in the process of collaging, unfortunately, I could not find all of their names after the final visual was created.

My work is supported by Trust-M project, funded by the Strategic Research Council of Finland (372404). 
\end{acks}


\bibliographystyle{ACM-Reference-Format}
\bibliography{chideology_paper}

@article{bodker_participatory_2018,
	title = {Participatory {Design} that {Matters}—{Facing} the {Big} {Issues}},
	volume = {25},
	issn = {1073-0516, 1557-7325},
	url = {https://dl.acm.org/doi/10.1145/3152421},
	doi = {10.1145/3152421},
	language = {en},
	number = {1},
	urldate = {2023-10-05},
	journal = {ACM Transactions on Computer-Human Interaction},
	author = {Bødker, Susanne and Kyng, Morten},
	month = feb,
	year = {2018},
	pages = {1--31},
}

@article{blessett_social_2019,
	title = {Social {Equity} in {Public} {Administration}: {A} {Call} to {Action}},
	issn = {2398-4910, 2398-4929},
	shorttitle = {Social {Equity} in {Public} {Administration}},
	url = {https://academic.oup.com/ppmg/advance-article/doi/10.1093/ppmgov/gvz016/5601191},
	doi = {10.1093/ppmgov/gvz016},
	language = {en},
	urldate = {2023-11-27},
	journal = {Perspectives on Public Management and Governance},
	author = {Blessett, Brandi and Dodge, Jennifer and Edmond, Beverly and Goerdel, Holly T and Gooden, Susan T and Headley, Andrea M and Riccucci, Norma M and Williams, Brian N},
	month = oct,
	year = {2019},
	pages = {gvz016},
}

@article{huybrechts_institutioning_2017,
	title = {Institutioning: {Participatory} {Design}, {Co}-{Design} and the public realm},
	volume = {13},
	issn = {1571-0882, 1745-3755},
	shorttitle = {Institutioning},
	url = {https://www.tandfonline.com/doi/full/10.1080/15710882.2017.1355006},
	doi = {10.1080/15710882.2017.1355006},
	language = {en},
	number = {3},
	urldate = {2024-01-24},
	journal = {CoDesign},
	author = {Huybrechts, Liesbeth and Benesch, Henric and Geib, Jon},
	month = jul,
	year = {2017},
	pages = {148--159},
}

@article{bodker_tying_2017,
	title = {Tying {Knots}: {Participatory} {Infrastructuring} at {Work}},
	volume = {26},
	issn = {1573-7551},
	shorttitle = {Tying {Knots}},
	url = {https://doi.org/10.1007/s10606-017-9268-y},
	doi = {10.1007/s10606-017-9268-y},
	language = {en},
	number = {1},
	urldate = {2024-05-20},
	journal = {Computer Supported Cooperative Work (CSCW)},
	author = {Bødker, Susanne and Dindler, Christian and Iversen, Ole Sejer},
	month = apr,
	year = {2017},
	pages = {245--273},
}

@book{frederickson_social_2010,
	address = {Armonk, N.Y},
	title = {Social equity and public administration: origins, developments, and applications},
	isbn = {978-0-7656-2471-0 978-0-7656-2472-7},
	shorttitle = {Social equity and public administration},
	language = {en},
	publisher = {M.E. Sharpe, Inc},
	author = {Frederickson, H. George},
	year = {2010},
	note = {OCLC: ocn427757246},
}

@inproceedings{henriques_feminist_2025,
	address = {New York, NY, USA},
	series = {{CHI} '25},
	title = {A {Feminist} {Care} {Ethics} {Toolkit} for {Community}-{Based} {Design}: {Bridging} {Theory} and {Practice}},
	isbn = {979-8-4007-1394-1},
	shorttitle = {A {Feminist} {Care} {Ethics} {Toolkit} for {Community}-{Based} {Design}},
	url = {https://dl.acm.org/doi/10.1145/3706598.3713950},
	doi = {10.1145/3706598.3713950},
	urldate = {2025-06-24},
	booktitle = {Proceedings of the 2025 {CHI} {Conference} on {Human} {Factors} in {Computing} {Systems}},
	publisher = {Association for Computing Machinery},
	author = {Henriques, Ana O and Carter, Anna R. L. and Severes, Beatriz and Talhouk, Reem and Strohmayer, Angelika and Pires, Ana Cristina and Gray, Colin M. and Montague, Kyle and Nicolau, Hugo},
	month = apr,
	year = {2025},
	pages = {1--26},
}

@inproceedings{christensen_researchers_2024,
	address = {Sibu Malaysia},
	title = {The {Researcher}'s {Plight}: {Guilt} and {Shame} in {Participatory} {Design} and {Action} {Research}},
	isbn = {979-8-4007-0654-7},
	shorttitle = {The {Researcher}'s {Plight}},
	url = {https://dl.acm.org/doi/10.1145/3661455.3669865},
	doi = {10.1145/3661455.3669865},
	language = {en},
	urldate = {2025-06-26},
	booktitle = {Participatory {Design} {Conference} 2024},
	publisher = {ACM},
	author = {Christensen, Lars Rune and Ahsan, Hasib},
	month = aug,
	year = {2024},
	pages = {19--23},
}

@book{tronto_moral_2020,
	edition = {1},
	title = {Moral {Boundaries}: {A} {Political} {Argument} for an {Ethic} of {Care}},
	isbn = {978-1-003-07067-2},
	shorttitle = {Moral {Boundaries}},
	url = {https://www.taylorfrancis.com/books/9781000107777},
	doi = {10.4324/9781003070672},
	language = {en},
	urldate = {2025-07-18},
	publisher = {Routledge},
	author = {Tronto, Joan C.},
	month = jul,
	year = {2020},
}

@book{ahmed_cultural_2014,
	address = {Edinburgh},
	edition = {Second edition},
	title = {The cultural politics of emotion},
	isbn = {978-0-7486-9114-2},
	url = {https://www.jstor.org/stable/10.3366/j.ctt1g09x4q},
	language = {eng},
	publisher = {Edinburgh University Press},
	author = {Ahmed, Sara},
	year = {2014},
}

@inproceedings{smith_contemporary_2025,
	address = {Aarhus N Denmark},
	title = {Contemporary {Participatory} {Design}: {Research} {Agendas} for {Societal} {Crisis}},
	isbn = {979-8-4007-2003-1},
	shorttitle = {Contemporary {Participatory} {Design}},
	url = {https://dl.acm.org/doi/10.1145/3744169.3744183},
	doi = {10.1145/3744169.3744183},
	language = {en},
	urldate = {2025-09-08},
	booktitle = {Proceedings of the sixth decennial {Aarhus} conference: {Computing} {X} {Crisis}},
	publisher = {ACM},
	author = {Smith, Rachel Charlotte and Huybrechts, Liesbeth and Simonsen, Jesper and Loi, Daria},
	month = aug,
	year = {2025},
	pages = {182--201},
}

@book{puig_de_la_bellacasa_matters_2017,
	title = {Matters of {Care}},
	isbn = {978-1-4529-5346-5},
	language = {eng},
	author = {Puig de la Bellacasa, María.},
	year = {2017},
	note = {OCLC: 1480916987},
}

@article{bartels_debate_2025,
	title = {Debate: {A} relational agenda for changing public administration research and practice},
	volume = {45},
	issn = {0954-0962, 1467-9302},
	shorttitle = {Debate},
	url = {https://www.tandfonline.com/doi/full/10.1080/09540962.2024.2402873},
	doi = {10.1080/09540962.2024.2402873},
	language = {en},
	number = {1},
	urldate = {2026-01-14},
	journal = {Public Money \& Management},
	author = {Bartels, Koen P. R. and Von Heimburg, Dina and Jordan, Gerald and Ness, Ottar},
	month = jan,
	year = {2025},
	pages = {3--5},
}

@inproceedings{petterson_expanding_2025,
	address = {Toronto ON Canada},
	title = {Expanding {Care} {Conceptualizations}: {An} {Integrative} {Literature} {Review} of {Care} in {HCI}},
	isbn = {979-8-4007-1484-9},
	shorttitle = {Expanding {Care} {Conceptualizations}},
	url = {https://dl.acm.org/doi/10.1145/3715335.3735486},
	doi = {10.1145/3715335.3735486},
	language = {en},
	urldate = {2026-01-23},
	booktitle = {Proceedings of the {ACM} {SIGCAS}/{SIGCHI} {Conference} on {Computing} and {Sustainable} {Societies}},
	publisher = {ACM},
	author = {Petterson, Adrian and Mattka, Jocelyn and Chandra, Priyank},
	month = jul,
	year = {2025},
	pages = {481--503},
}

@article{vaughan_pieced_2005,
	title = {Pieced {Together}: {Collage} as an {Artist}'s {Method} for {Interdisciplinary} {Research}},
	volume = {4},
	copyright = {https://journals.sagepub.com/page/policies/text-and-data-mining-license},
	issn = {1609-4069, 1609-4069},
	shorttitle = {Pieced {Together}},
	url = {https://journals.sagepub.com/doi/10.1177/160940690500400103},
	doi = {10.1177/160940690500400103},
	language = {en},
	number = {1},
	urldate = {2026-02-04},
	journal = {International Journal of Qualitative Methods},
	author = {Vaughan, Kathleen},
	month = mar,
	year = {2005},
	pages = {27--52},
}


\end{document}